\documentclass[12pt]{iopart}
\usepackage{iopams} 
\usepackage{graphicx}
\usepackage{bm}
\usepackage{color} 

\begin{document}

\title[Interplanar magnetic exchange in CoPS$_3$]{Interplanar magnetic exchange  in CoPS$_3$}

\author{A. R. Wildes$^{1*}$, B. F{\aa}k$^1$, U. B. Hansen$^1$, A. Ivanov$^1$, M. Enderle$^1$, L. Puertas Pel{\'a}ez$^{1,2}$}
\ead{wildes@ill.fr}

\address{$^1$Institut Laue-Langevin ,71 avenue des Martyrs, CS 20156, 38042 Grenoble Cedex 9,  France}
\address{$^2$Universidad Carlos III de Madrid Av. de la Universidad, 30, 28911 Legan{\'e}s, Madrid, Spain.}


\begin{abstract}
Neutron three-axis spectrometry has been used to determine the interplanar magnetic exchange parameter  in the magnetic van der Waals compound CoPS$_3$.  The exchange is found to be small and antiferromagnetic, estimated to be $0.020 \pm 0.001$ meV, which is surprising considering that the magnetic structure is correlated ferromagnetically between the $ab$ planes.  A possible explanation, involving a small anisotropy in the exchanges, is proposed. The results are discussed with reference to the other members of the transition metal-PS$_3$ compounds.
\end{abstract}

%
\vspace{2pc}

%
%
%
%

\section{Introduction}

Layered van der Waals compounds with intrinsic magnetic properties have become a popular subject for research following the discovery of graphene and the realization that other van der Waals compounds can be delaminated down to a monolayer \cite{Park, Wang}.  The compounds offer the possibility for deep insight into the fundamental aspects of magnetism in a true two-dimensional limit, and have the potential for use for a wide variety of applications in graphene-inspired technology.   

Both motivations require the development of an appropriate Hamiltonian for the quantum energies associated with the magnetism in the compounds, allowing the properties to be comprehensively understood and predicted.  Modern computational methods like density functional theory are powerful tools in this regard, but the results must be coherent with those from experiments.  Data from neutron scattering experiments are particularly relevant as they provide a quantitative measurement of the dynamic structure factor, $S\left({\bf{Q}},\omega\right)$, and a direct test of the Hamiltonian.  The data are limited by statistics due to the relatively weak brightness of even the highest flux neutron sources, meaning experiments usually focus on bulk samples.  However, the interplanar magnetic exchanges in layered van der Waals compounds tend to be very small meaning that even bulk samples are highly two-dimensional.  Furthermore, the data provide a solid benchmark to test computations before progressing to a monolayer.

\begin{figure}
  \includegraphics[scale=0.5]{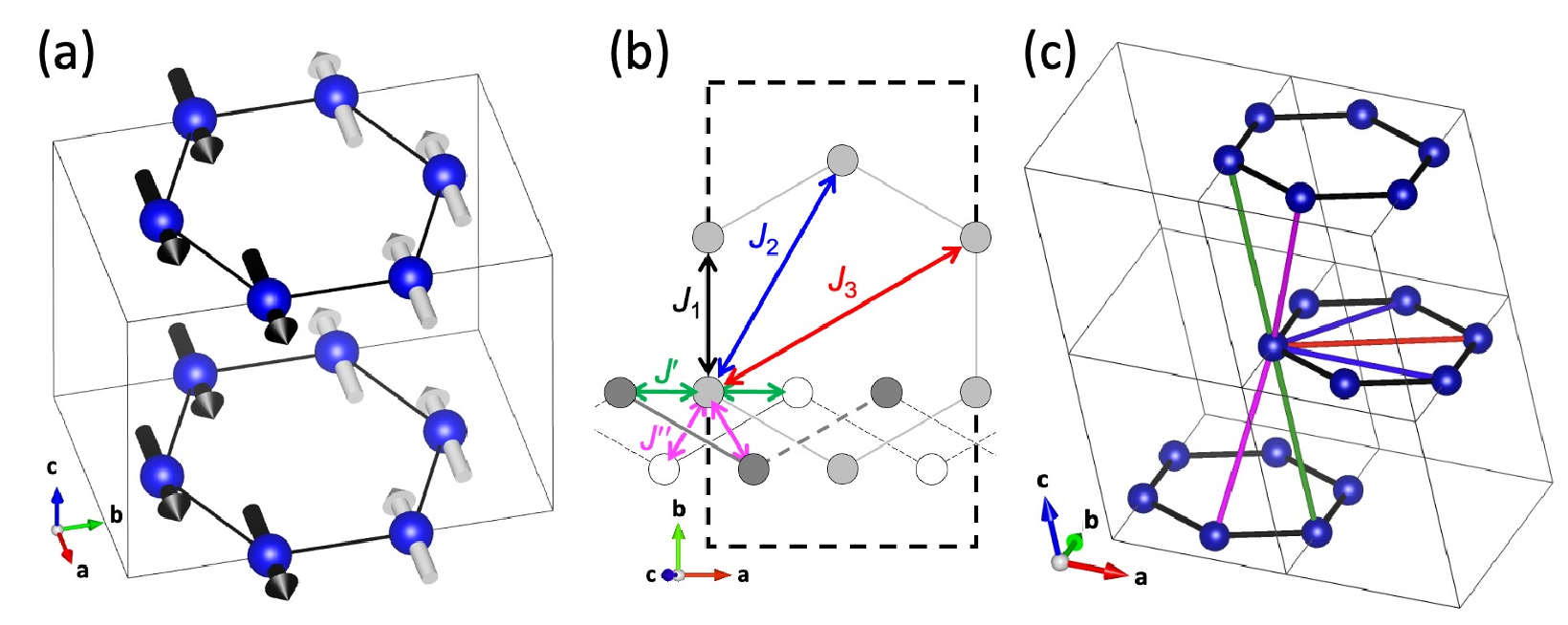}
  \caption{\label{fig:MagStruc} Magnetic structure for CoPS$_3$.  (a) Isometric view of the unit cell. (b) View along the {\bf c}$^*$ axis showing the exchange parameter pathways.  The Co$^{2+}$ ions in a plane are shown in light grey.  Ions one plane below are shown in white, and ions one plane above are shown in dark grey.  Intraplanar exchanges between first, second and third nearest neighbours are shown as $J_1$ (black), $J_2$ (blue) and $J_3$ (red) respectively.  Interplanar exchanges are shown as $J^{\prime}$ (green) and $J^{\prime\prime}$ (magenta). (c) Isometric view of (b) showing the exchange parameters using the same colour code.  Figures (a) and (c) created using the VESTA software package \cite{VESTA}.}
\end{figure}

CoPS$_3$ belongs to the TM-PS$_3$ (TM = Mn, Fe, Co, Ni) family of magnetic van der Waals compounds.   The family has been extensively studied for many years \cite{Brec}.  All have the monoclinic space group $C\frac{2}{m}$ and the magnetic ions form a honeycomb lattice in the $ab$ planes.  Their magnetic structures are collinear antiferromagnets with the moment axes constrained to lie in the $ac$ plane.   CoPS$_3$ is the least studied of the family, most likely due to it being the most difficult to synthesise.  Its magnetic structure, shown in figure \ref{fig:MagStruc}(a), consists of ferromagnetic zig-zag chains running parallel to the ${\bf{a}}$ axis that are antiferromagnetically correlated along the ${\bf{b}}$ axis \cite{CoPS_elastic}.  The moments are all approximately collinear with the ${\bf{a}}$ axis.

Neutron inelastic scattering has been used to measure the spin wave dispersion for all the compounds in the family \cite{Lancon,Wildes98,Wildes22,Sheie,CoPS_inelastic}.  The data could be accurately modelled using linear spin wave theory with the Hamiltonian:
\begin{equation}
  \begin{array}{ll}
  \hat{\mathcal{H}}=&\frac{1}{2}\sum_{\left< ij \right>}J_{ij} \left(S^x_iS^x_j + S^y_iS^y_j + S^z_iS^z_j \right) \\
  &+ D^x\sum_i\left[\left(S^{x}_i\right)^2-\left(S^{y}_i\right)^2\right] + D^z\sum_i\left(S^{z}_i\right)^2,
  \label{eq:Hamiltonian_Wildes}
  \end{array}
\end{equation}
where $J_{ij}$ is the magnetic exchange interaction between moments ${\bf S}_i$ and ${\bf S}_j$, $D^x$ and $D^z$ are single-ion anisotropy terms with orthogonal axes.  The $z$ axis was normal to the $ab$ planes, and $y$ was parallel to the ${\bf{b}}$ axis.  The Hamiltonian is defined such that negative $J$ indicates a ferromagnetic exchange, negative $D$ indicates a uniaxial anisotropy and positive $D$ indicates a planar anisotropy.  All the compounds could be fitted assuming a spin-only model, which is $S = 3/2$ for Co$^{2+}$ in the local octahedral environment with elogated trigonal distortion \cite{Winter}.  Exchanges up to the third nearest-neighbour in the $ab$ plane must be included to reproduce the measured neutron scattering from the spin waves.  The exchanges, labelled $J_1$, $J_2$ and $J_3$ for the first, second and third intraplanar neighbours respectively, are shown in figures \ref{fig:MagStruc}(b) and (c).  The values for CoPS$_3$ were determined to be $J_1 = -1.37\left(7\right)$ meV, $J_2 = 0.09\left(5\right)$ meV, $J_3 = 3.0\left(1\right)$ meV, $D^x = -0.77\left(3\right)$ meV, and $D^z = 6.07\left(4\right)$ meV \cite{CoPS_inelastic}.  The signs of the anisotropy agree with the local environment and the crystal electric field parameters estimated from the susceptibility \cite{CoPS_elastic}.

Bulk crystals of TM-PS$_3$ are good approximations of two-dimensional magnets, however they do have an interplanar exchange leading to a small spin wave dispersion parallel to the ${\bf{c}^*}$ axis.  To first approximation, each ion in a plane has four interplanar neighbours, with two approximately equidistance nearest-neighbours on each adjacent plane \cite{Ouvrard85}.  Examples are shown in figures \ref{fig:MagStruc}(b) and (c).  One pair of neighbours is displaced by one unit cell along the $c$-axis from the central ion, i.e. at $\pm{\bf{c}}$, and the relevant exchanges are labelled as $J^{\prime}$.  The other pair of neighbours is at $\pm\frac{1}{2}{\bf{a}} + \frac{1}{6}{\bf{b}} \pm {\bf{c}}$ or $\pm\frac{1}{2}{\bf{a}} - \frac{1}{6}{\bf{b}} \pm {\bf{c}}$, depending on the position of the ion in the central plane.  These symmetry-equivalent exchanges are labelled as $J^{\prime\prime}$.  Values for the interplanar exchanges have been determined from neutron scattering data for MnPS$_3$ \cite{Wildes98}, FePS$_3$ \cite{Lancon}, and NiPS$_3$ \cite{Wildes22,Sheie}, however the counting statistics and data quality were insufficient to quantify any interplanar dispersion in CoPS$_3$ \cite{CoPS_inelastic}.  A quantitative determination of this exchange is the last parameter required for a full characterisation of the spin waves in TM-PS$_3$ using neutron inelastic scattering.

This article follows the previous report on the intraplanar exchange and anisotropy parameters \cite{CoPS_inelastic}, now describing a dedicated experiment using neutron three-axis spectrometry to determine the interplanar magnetic exchange parameters in CoPS$_3$.  The technique has significant advantages for such a study over the time-of-flight neutron spectrometry measurements used to quantify the intraplanar parameters, having substantially more intensity for an individual neutron momentum, ${\bf{Q}}$, and energy, $\omega$, transfer and with a well-defined, quantitative, and analytical expression for the instrument resolution.   The data have again been analysed and interpreted using linear spin wave theory.  The results are discussed with reference to the interplanar exchanges determined for the other members of the family.

\section{Methods}
Neutron scattering experiments were performed using IN8, the thermal three-axis spectrometer at the Institut Laue-Langevin \cite{Piovano_IN8, IN8_DOI}.  The instrument was configured with a bent silicon $\left(1~1~1\right)$ monochromator and a doubly-focussed pyrolytic graphite $\left(0~0~2\right)$ analyser.  The analyser fixed the final neutron energy to 14.68 meV and higher-order wavelength contamination was suppressed using a pyrolytic graphite filter between sample and analyser.  Temperature control of the sample was achieved using a liquid helium cryostat.

Measurements were performed on the same composite sample used for the three-axis experiments in reference \cite{CoPS_inelastic} comprising of four co-aligned crystals glued to an aluminium plate.  The sample was realigned to measure optimally the spin wave dispersion normal to the layer planes.  The neutron cross-section is directly proportional to the magnetic form factor squared, which is has its maximum at $Q = 0$ for Co$^{2+}$.  Spin wave measurements should thus be performed at the smallest possible $Q$ to maximise the neutron intensity.  The sample was aligned with $\left(h~2h~l\right)$ as the nominal scattering plane.  For the chosen final neutron energy, kinematic conditions meant that the $1~2~0$ and $1~2~\overline{1}$ peaks were the Brillouin zone centres with the lowest $Q$ that could access energy transfers up to 20 meV.    The previous experiments showed that the lowest energy mode in the spin wave spectrum for CoPS$_3$ was $\approx 14.5$ meV \cite{CoPS_inelastic}, thus the dispersion of this mode along $l$ could be followed.

The data were analysed using linear spin wave theory in the same manner as that previously reported \cite{CoPS_inelastic}.  It was assumed in the analysis that there was no distinction between the two interplanar neighbours and that their exchange parameters were equal, i.e. $J^{\prime} = J^{\prime\prime}$.

 \section{Results}

\begin{figure}
  \includegraphics[scale=0.5]{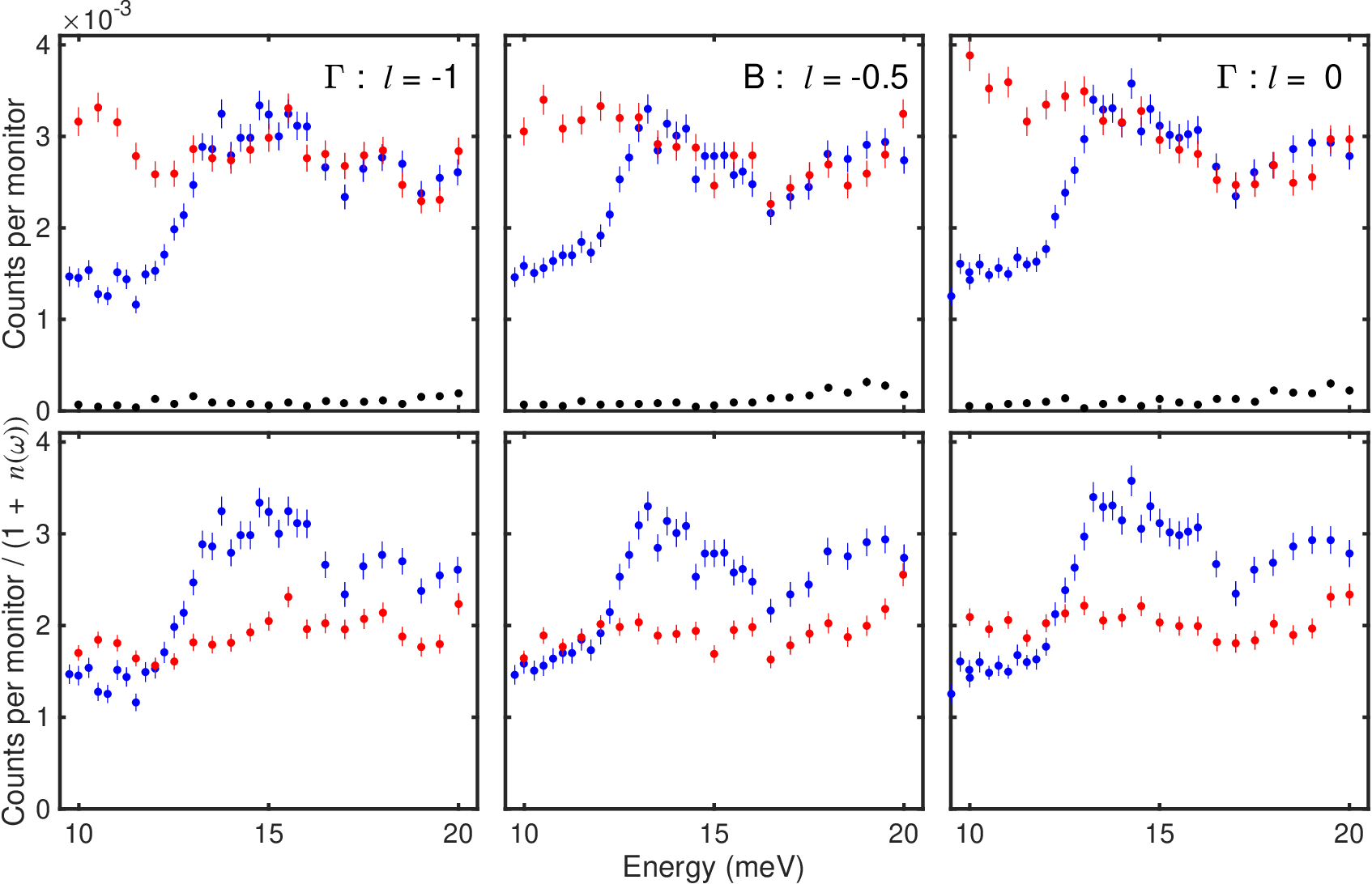}
  \caption{\label{fig:Tplots} Neutron three-axis data from CoPS$_3$, measured at $1~2~l$ for a selection of $l$.  The top row shows count rates for sample temperatures of 1.7 K (blue points) and 150 K (red points), along with similar scans when the analyser was rotated by $5^{\circ}$ (black points).  The lower row shows the same CoPS$_3$ data as the top row divided by the temperature factor, $1+n\left(\omega\right)$.}
\end{figure}

Scans at constant $\bf{Q}$, varying energy transfer, were performed at a series of $1~2~l$ positions for $-1 \le l \le 0$.  The measurements tracked the spin wave dispersion along the $\Gamma - B$ trajectory in the Brillouin zone, with $\Gamma$ being the Brillouin zone centres at $1~2~\overline{1}$ and $1~2~0$ and $B$ being the Brillouin zone boundary at $1~2~\frac{\overline{1}}{2}$.  Examples are shown in the top row of figure \ref{fig:Tplots}.  Repeated measurements performed with the analyser rotated by $5^{\circ}$ are also shown, demonstrating that the intrinsic background of the instrument was very low.  

Peaks are observed at $\approx 14$ meV when the sample temperature was 1.7 K, consistent with the lowest energy spin wave mode.  The peaks appear on a large background.  The scattering at the lowest energy transfers, $\lesssim 12$ meV, is weaker than that at larger energy transfers, $\gtrsim 17$ meV, consistent with the previous measurements of the same sample with a different scattering plane on IN8 \cite{CoPS_inelastic}.  Neutron time-of-flight measurements showed only one spin wave branch in this energy range \cite{CoPS_inelastic}, establishing that the scattering above 17 meV was spurious and not due to spin waves.     

The three-axis background was previously taken to be largely phononic, coming from the sample mount.  Measurements were thus performed above the N{\'e}el temperature, at 150 K, in an attempt to  quantify and subtract the background.  The results are shown in figure \ref{fig:Tplots}.  The top row shows a superposition of count rates measured at 1.7 K and 150 K.  The scattering above $\sim13.5$ meV is essentially independent of temperature.  The lower row of the figure shows the count rates divided by the temperature factor, $1+n\left(\omega\right)$, based on Bose statistics.  If the background were purely phononic and the phonon energies were temperature-independent, a subtraction would result in a clean peak.  The 150 K data are slightly greater than the 1.7 K data at the lowest energies, and slightly smaller at the highest energies.  The observations are consistent with the background being due to a combination of a temperature-dependent phononic part from the sample mounting, and a temperature-independent part from the sample environment, which was revealed in a limited number of measurements of an empty cryostat post-experiment.  Consequently, no attempt was made to subtract the background which was instead modelled using a sloped, linear function.

The data were fitted with the RESCAL5 library \cite{RESCAL} for MATLAB, using a function that included a four-dimensional convolution with the instrument resolution.  The function accounted for the local curvature of the dispersion surface, following the equation:
\begin{equation}
  \label{eq:LocalDisp}
   E_q = 240 h^2 + 75 k^2 + E_l,
 \end{equation}
where energies are in meV.  The equation is consistent with the dispersion calculated using the exchange parameters previously published for CoPS$_3$, which showed that the dispersion between the $\Gamma$ and $B$ points in the Brillouin zone is locally parabolic in $h$ and $k$ and dispersionless along $l$.  The data were initially individually fitted with the background slope and intercept as free parameters.  Inspection of the fitted background parameters showed a variation of less than 10\%.  The parameters were subsequently averaged and the data refitted with a common, fixed background function.  The only free parameters in the final fits were an amplitude to scale the spin wave peak intensity and  $E_l$, the local energy at $1~2~l$.  Fits with free and fixed background parameters gave qualitatively the same trends for these parameters.

\begin{figure}
  \includegraphics[scale=1]{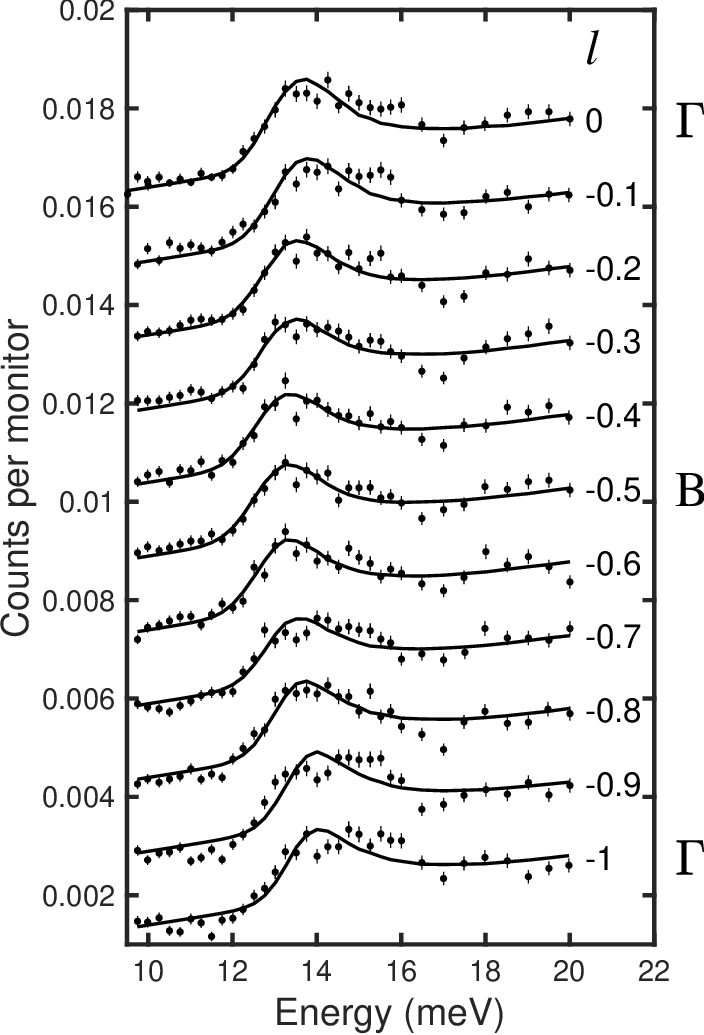}
  \caption{\label{fig:Waterfall} Counts for monitor of the inelastic scattering along $1~2~l$ for CoPS$_3$.  The values for $l$ are shown in the plot, and the data have been shifted vertically by 0.0015 for clarity.  Fits of equation \ref{eq:LocalDisp}, convoluted with the instrument resolution and including a linear background, are also shown for each data set.}
\end{figure}

Figure \ref{fig:Waterfall} shows the data together with the final fits.  The data are shifted vertically for clarity.  Close inspection shows that the peaks are slightly asymmetric, with a tail at higher energy transfers due to the curvature of the local dispersion surface.

\begin{figure}
  \includegraphics[scale=0.5]{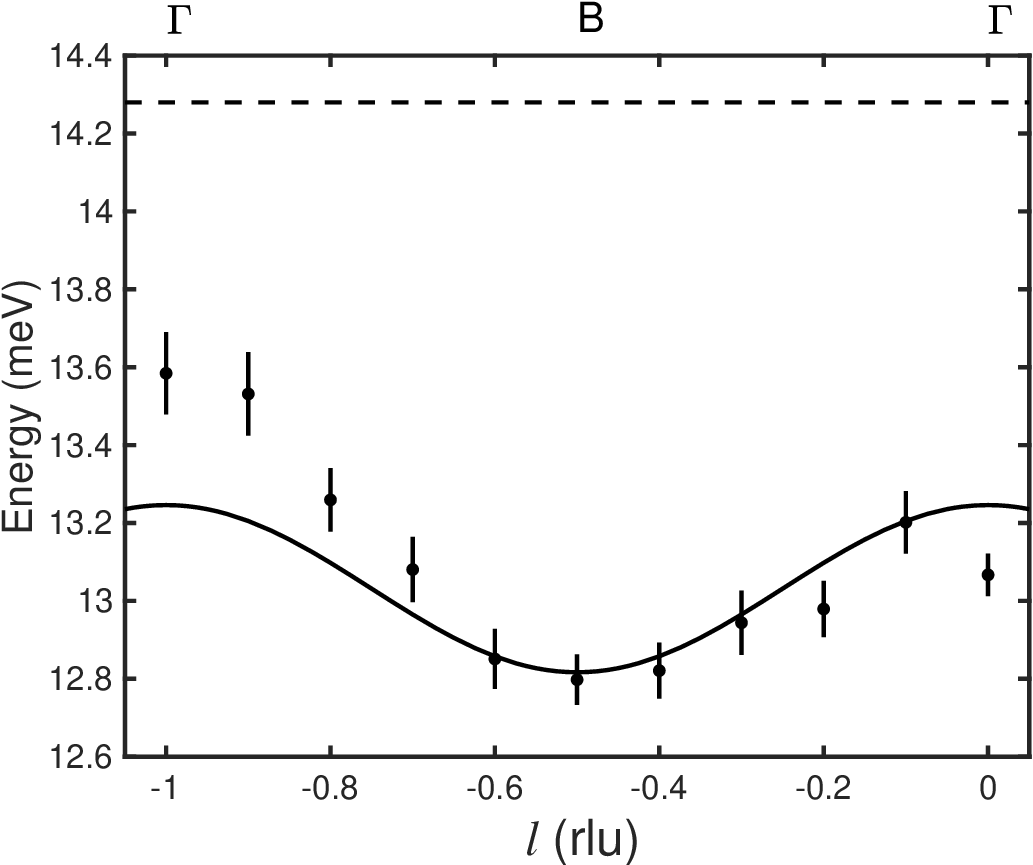}
  \caption{\label{fig:Dispersion} Estimates for the energies of the interplanar spin waves as a function of $l$.  The black points mark the values for $E_l$  determined from the fits shown in figure \ref{fig:Waterfall}.  The fit of equation \ref{eq:Ldisp} to $E_l$ is shown as a black line.  The spin wave energy at the Brillouin zone centre determined from analysis of the previous data \cite{CoPS_inelastic} is shown as a black dashed line.}
\end{figure}

The fitted values for $E_l$ are shown in figure \ref{fig:Dispersion}, along with the value for the spin wave energy at the $\Gamma$ point calculated using the published exchange parameters and anisotropies \cite{CoPS_inelastic}.  The values from the current analysis are slightly lower in energy compared to the previous estimate, reflecting the improved statistics of the new data and the better description of the resolution.  The new energies can be reproduced if the magnitudes of the published parameters are reduced by $\approx 5\%$, which is close to their uncertainties.

The new data show a small dispersion with a minimum at the Brillouin zone boundary.  The result is surprising as this indicates that the interplanar exchange in CoPS$_3$ is weakly antiferromagnetic despite the planes being ferromagnetically correlated in the magnetic structure.  

As previously shown \cite{CoPS_inelastic}, the spin waves on CoPS$_3$ are well-described using equation \ref{eq:Hamiltonian_Wildes} with two large single-ion anisotropies and isotropic exchange parameters.  The dispersion along the $\Gamma-B$ direction calculated using equation \ref{eq:Hamiltonian_Wildes} follows a cosine function and energies at the high-symmetry points are given by:
\begin{equation}
  \begin{array}{ll}
  \Gamma \equiv 0 0 0: & 2S\left(2D^x\left(D^x-D^z-J_1-4J_2-3J_3\right)\right)^{\frac{1}{2}}, \\
  B \equiv 0 0 {\frac{1}{2}}: & 2S\left(2\left(D^x+J^\prime+J^{\prime\prime}\right)\left(D^x-D^z-J_1-4J_2-3J_3+2J^\prime+2J^{\prime\prime}\right)\right)^{\frac{1}{2}}.
  \end{array}
\end{equation}
The values for $E_l$ in figure \ref{fig:Dispersion} were therefore fitted with the equation:
\begin{equation}
  \label{eq:Ldisp}
  E_l = M + N\cos\left(2\pi l\right),
\end{equation}
and the fit is also shown in figure \ref{fig:Dispersion}.  

The fitted parameters of $M = 13.03 \pm 0.05$ meV and $N = 0.21 \pm 0.07$ meV give spin wave energies of 13.25 meV and 12.82 meV at the $l = \overline{1},0$ and $l=\overline{\frac{1}{2}}$ positions respectively.  The values for $M$ and $N$ and the previously-published exchange parameters and anisotropies were used to calculate an interplanar exchange of $J^{\prime} = J^{\prime\prime} =0.020 \pm 0.001$.  The value remains within the uncertainty if the anisotropies and intraplanar exchange parameters are scaled to match the spin wave energy given by equation \ref{eq:Ldisp} at the $\Gamma$ point.

\section{Discussion}
The experimental observation that the interplanar exchange in CoPS$_3$ is antiferromagnetic is surprising given that the nearest neighbours between $ab$ planes are ferromagnetically correlated.  The observation warrants further discussion, testing the assumptions inherent in the modelling of the data and including a comparison with the other members of the TM-PS$_3$ family.

The magnetic structures in the $ab$ planes for CoPS$_3$, FePS$_3$ and NiPS$_3$ are qualitatively identical, having the same zig-zag chains parallel to $\bf{a}$.  They differ in the direction of the ordered magnetic moments, being almost parallel to the $\bf{a}$ axis for CoPS$_3$ and NiPS$_3$ and normal to the $ab$ planes for FePS$_3$.  The planes are ferromagnetically correlated for CoPS$_3$ and NiPS$_3$ and antiferromagnetically correlated for FePS$_3$ meaning that, in the limit of small interplanar exchange, it is the sum $J^{\prime} + J^{\prime\prime}$ that is relevant when diagonalizing equation \ref{eq:Hamiltonian_Wildes} for these compounds.  Consequently, it may be more accurate to state that the current analysis determines the interplanar exchange in CoPS$_3$ to be $J^{\prime} + J^{\prime\prime} = 0.040 \pm 0.002$ meV.

The in-plane magnetic structure for MnPS$_3$ differs significantly from its sister compounds in that it has a N{\'e}el-type antiferromagnetic order, with ferromagnetic correlations between the planes.  This means that each moment is ferromagnetically correlated to its  $J^{\prime}$ neighbours and antiferromagnetically correlated to its $J^{\prime\prime}$ neighbours.  The important parameter describing the interplanar dispersion is $J^{\prime} - J^{\prime\prime}$.  The interplanar dispersion for MnPS$_3$ is clearly ferromagnetic \cite{Wildes98}, meaning $J^{\prime} - J^{\prime\prime} < 0$ and hence $J^{\prime} \ne J^{\prime\prime}$.  Certainly, the MnPS$_3$ spin wave dispersion was analysed with a naive assumption that $J^{\prime\prime} = 0$ \cite{Wildes98}.  The comparison questions whether and why $J^{\prime}$ and $J^{\prime\prime}$ can differ in TM-PS$_3$.    

Close examination of the crystal structures \cite{Ouvrard85} to determine possible superexchange pathways does not immediately suggest any reason for a difference.  As noted in the introduction, all four members of the family have the same monoclinic space group $C\frac{2}{m}$.  Their structures are slightly distorted from having a hexagonal space group, and in the case of FePS$_3$ and CoPS$_3$ the distortion is so small that the Bragg peaks can be indexed perfectly using a hexagonal unit cell \cite{Brec, Ouvrard85}.  Nevertheless, the interplanar exchanges are small, especially for CoPS$_3$, FePS$_3$ and MnPS$_3$, and subtle structural differences in bond angles and lengths could result in small differences between $J^{\prime}$ and $J^{\prime\prime}$, even potentially involving a change in sign between them.

\begin{figure}
  \includegraphics[scale=0.5]{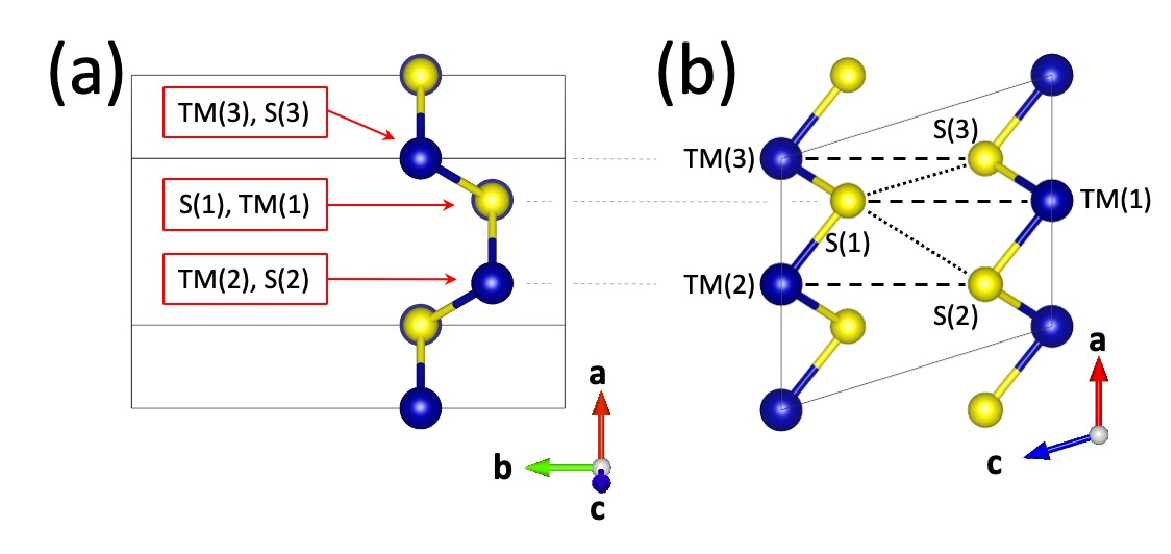}
  \caption{\label{fig:Interplanar}  Projections of the unit cell for TM-PS$_3$ with metal atoms (TM) forming zigzag chains parallel to the ${\bf{a}}$ axis, and the sulfur atoms (S) that likely mediate the interplanar magnetic exchange.  The unit cell is viewed (a) along the $\bf{c^*}$ axis and (b) along the $-{\bf{b}}$ axis.  Three TM atoms and three S atoms are labelled to show examples of interplanar superexchange pathways, with the potential pathways indicated by dashed lines in (b).  Figure created using the VESTA software package \cite{VESTA}}.
\end{figure}

The magnetic superexchange is most likely to be mediated through sulfur atoms.  Figure \ref{fig:Interplanar} shows the $C\frac{2}{m}$ unit cell for TM-PS$_3$.  Selected atoms to illustrate possible interplanar exchange pathways are included, with TM(n) designating transition metal atoms and S(n) denoting sulfur.  The $J^{\prime}$ exchange couples the atoms TM(1) and TM(2).  Two suggested superexchange pathways involve a single sulfur, TM(1)-S(1)-TM(2) or TM(1)-S(2)-TM(2), with bond angles $\sim128.5^{\circ}$.  A second involves a coplanar double sulfur route, TM(1)-S(2)-S(1)-TM(2), with each bond angle $\sim97^{\circ}$.  The $J^{\prime\prime}$ exchange couples the atoms TM(1) and TM(3), again having single sulfur and double sulfur routes via S(1) and S(3).  The bond angles and inter-atomic distances for similar $J^{\prime}$ and $J^{\prime\prime}$-pathways vary only by a few tenths of a percent  for a given TM-PS$_3$ compound, and by a few percent between different family members for equivalent pathways.  However, the small differences may be enough to influence the sign and magnitude of the exchanges, especially if the single- and double-sulfur pathways are in competition.  The influence is likely to be greater in MnPS$_3$ and NiPS$_3$ whose crystal structures are more distorted from a perfect hexagonal structure than CoPS$_3$ and FePS$_3$ \cite{Brec, Ouvrard85}.

Examination of the superexchange pathways, however, will not immediately reconcile the antiferromagnetic exchange, $\left(J^{\prime} + J^{\prime\prime}\right) > 0$, in CoPS$_3$ with its magnetic structure if equation \ref{eq:Hamiltonian_Wildes} is rigorously applied.  The ground state energy for the magnetic structure of CoPS$_3$ can be calculated from equation \ref{eq:Hamiltonian_Wildes} in the mean-field approximation to give:
\begin{equation}
  \begin{array}{ll}
  E_{\rm GS}=&S^2 \left(2J_1-4J_2-6J_3\right) \\
  &+4S^2\left(J^{\prime} + J^{\prime\prime}\right) \\
  &+4\left(D_z [S^{z}]^2+D_x([S^{x}]^2-[S^{y}]^2) \right)  \\
  \label{eq:GS_1}
  \end{array}
\end{equation}
where $ E_{\rm GS}$ is the energy per unit cell.  Large single-ion anisotropies, as are present in CoPS$_3$, will favour a moment axis but will not influence the ground state energy so long as the exchange parameters, $J$, are isotropic. With this assumption, an antiferromagnetic interplanar exchange would result in antiferromagnetic coupling between the planes.

A possible solution presents itself if the interplanar exchange parameters, $J^{\prime}$ and $J^{\prime\prime}$ are anisotropic.  Equation \ref{eq:GS_1} may then be rewritten:
\begin{equation}
  \begin{array}{ll}
  E_{\rm GS}=&S^2 \left(2J_1-4J_2-6J_3\right) \\ 
  &+4 \left\{\left(J^{\prime}_{xx}+J^{\prime\prime}_{xx}\right)\left[S^{x}\right]^2 + \left(J^{\prime}_{yy}+J^{\prime\prime}_{yy}\right)\left[S^{y}\right]^2 + \left(J^{\prime}_{zz}+J^{\prime\prime}_{zz}\right)\left[S^{z}\right]^2\right\}\\
  &+4\left(D_z [S^{z}]^2+D_x([S^{x}]^2-[S^{y}]^2) \right).  \\
  \label{eq:GS_3}
  \end{array}
\end{equation}
In the case of CoPS$_3$, $D^x$ and $D^z$ are large compared to all the exchange parameters and favour the alignment of the moments parallel to the $x$-axis.  The ground state structure depends only on $\left[S^x\right]^2$ and the sign of $\left(J^{\prime}_{xx}+J^{\prime\prime}_{xx}\right)$ determines whether the moments are ferro- or antiferromagnetically coupled.  Even a small $\left(J^{\prime}_{xx}+J^{\prime\prime}_{xx}\right) < 0$ would favour ferromagnetic interplanar correlations.  The interplanar dispersion, however, is governed by the transverse spin components, $S^y$ and $S^z$.  The dispersion would appear to be antiferromagnetic if $\left(J^{\prime}_{yy}+J^{\prime\prime}_{yy}\right)$ and $\left(J^{\prime}_{zz}+J^{\prime\prime}_{zz}\right)$ were both positive.

A similar discussion may be held for FePS$_3$, where the strong anisotropy axis is along $z$.  In this case, the sign of $\left(J^{\prime}_{zz}+J^{\prime\prime}_{zz}\right)$ dictates the ground state.  The planes are antiferromagnetically coupled in this compound hence the sum must again be positive, consistent with CoPS$_3$.  The interplanar dispersion in FePS$_3$ is determined by the $J_{xx}$ and $J_{yy}$ terms in equation \ref{eq:GS_3} and is again antiferromagnetic \cite{Lancon}, leading to the speculation that $\left(J^{\prime}_{yy}+J^{\prime\prime}_{yy}\right) > 0$ in both compounds.  The third component,  $\left(J^{\prime}_{xx}+J^{\prime\prime}_{xx}\right)$, can be either sign in FePS$_3$ so long as the net sum of the $x$ and $y$ components is positive. 

The quality of the CoPS$_3$ data is insufficient to test for any anisotropy in the interplanar exchanges, and the discussion serves only to propose a plausible explanation for the experimental observation.  The explanation does not contradict the arguments in \cite{CoPS_inelastic} which concluded that, to first order, the anisotropy in CoPS$_3$ is single-ion-like and the exchanges are isotropic.  The proposition involves a small anisotropy from very different one-sulfur and two-sulfur superexchange pathways on small interplanar exchanges, serving as a higher-order perturbation.  The proposition also does not constrain $J_{\alpha\alpha}^{\prime}$ and $J_{\alpha\alpha}^{\prime\prime}$ to be equal, or even to have the same sign, as it is their sum that matters.

Computation studies will be useful to determine the underlying physics behind the values for the exchange parameters.  There have been numerous attempts to reproduce the magnetic exchange parameters of TM-PS$_3$ using first principles calculations \cite{Chittari, Koo, Olsen, Yan, Amirabbasi}.  Most of the calculations treat the compounds in the two-dimensional limit.  There have been some successes to reproduce the measured spin wave spectra from neutron scattering, particularly for MnPS$_3$ and NiPS$_3$, but the calculations struggle when applied to the high-anisotropy compounds FePS$_3$ and CoPS$_3$.  It must be reinforced that neutron scattering provides a quantitative measurement of the dynamic structure factor, $S\left({\bf{Q}},\omega\right)$, and a direct test of the Hamiltonian.  The published neutron scattering results  for TM-PS$_3$ compounds are accurate measures of their spin wave spectra. Future calculations must reproduce the spin wave energies and neutron intensities to be representative of the correct physics driving the magnetic properties.

\section{Conclusions}
The interplanar exchange for CoPS$_3$ has been determined from neutron spectrometry to be weak and antiferromagnetic.  The determination is surprising given the ferromagnetic interplanar correlation of the magnetic structure, however could be made plausible by the presence of different competing contributions to the interplanar exchange and a tiny exchange anisotropy on those that collaborates with the strong single-ion anisotropy. This assumption would provide a coherent explanation both for CoPS$_3$ and for FePS$_3$ which is also a magnetically anisotropic compound that has an almost identical crystal structure, N{\'e}el temperature, and magnetic structure albeit with a different moment direction.  

\ack
We thank the Institut Laue-Langevin, proposal number 4-01-1803 \cite{IN8_DOI} for the allocation of beam time, and F. Charpenay for technical assistance during the experiment.  ARW thanks Dr. Duc Le for assistance with spin wave calculations.


\end{document}